# Thin films as practical quantum materials: a status quo and beyond

*Chaehyeong Ha and Yoon Jang Chung\**

Department of Chemical and Biological Engineering, Korea University, Seoul 02841, Republic of Korea

\*Corresponding author (E-mail: yoonjang@korea.ac.kr)

**Abstract**

Quantum materials have been in the limelight for several years now. These materials exhibit intriguing quantum phenomena, which when harnessed properly, promise extraordinary advancements across various scientific and technological domains. To fully exploit their potential, it is imperative to synthesize such quantum materials in thin film form so that they are compatible with well-established device fabrication techniques. In this perspective, an overview of the current status and future directions of thin film quantum material synthesis is provided. The criteria for quantum materials are discussed, as well as the many benefits of preparing them as thin films. Prominent deposition techniques such as molecular beam epitaxy and chemical vapor deposition are reviewed along with potential contenders. Despite challenges, progress in thin film quantum material technology holds the potential to realize practical devices with unprecedented functionalities.

**Introduction**

Breakthroughs in materials often revolutionize human society. As such, the historic timelines of mankind are commonly categorized with respect to the primary material utilized during a specific era[1]. The discovery of metals granted access to substantially sturdier tools compared to before, and the life of a steel-age man was radically different from that of a stone-age man. More recently, the advent of semiconductors transformed our everyday lives in previously inconceivable ways through inventions like the solid-state transistor[2-4].

An intriguing question to ask is what comes next. Many candidates may exist, but perhaps at the forefront of innovation are quantum materials[5,6]. Such systems exhibit intriguing emergent quantum phenomena such as superconductivity[7-9], topological band structure[10-12], or interaction-driven quasiparticle[13-15] and spin texture formation[16-18], which open up unprecedented possibilities across a wide range of scientific disciplines. Naturally, we are prompted to imagine the new opportunities that will appear as quantum material technologies develop and mature.

Certain prerequisites must be met for a quantum material to fully display its unusual properties. Since the electronic characteristics of quantum materials typically derive from convoluted interactions between lattice, spin, orbit, and charge subband degrees of freedom, even minute perturbations to the system can destroy the intricate balance necessary for their observation. Usually, low-temperature environments are preferred, and materials in which scattering events can be greatly suppressed are imperative. Sure enough, numerous growth techniques have been implemented for the synthesis of high-quality, single-crystal quantum materials following their progressive rise in prominence[19-21].

Perhaps the most prevalent are methods that can yield millimeter-size bulk crystals, such as vapor phase transport or flux growth[22,23]. The versatility of these approaches allow the relatively straightforward fabrication of a plethora of quantum materials, which are oftentimes complex multi-element compounds[21, 24-26]. Indeed, there have been several reports regarding experimental investigations of strongly-correlated electron phases or non-trivial band topologies in condensed matter systems realized by bulk single crystal growth[27-29].

Alternatively, thin film deposition is another popular quantum material growth procedure. This is because thin film structures grown on commercial substrates are more compatible with device fabrication schemes such as photolithography compared to the faceted surfaces of unconventional bulk crystals. Thin film processes are usually easier to scale up as well. In a sense then, to exploit quantum materials in a practical manner, it is crucial to have the capability to prepare them in thin film form. In this perspective, we review the current status of thin film quantum materials and discuss future research directions.

**What is quantum or not**

It is not so trivial to establish an unambiguous definition for what exactly a quantum material is. While a number of different interpretations exist, the description given at a recent workshop hosted by the United States Department of Energy captures the essence: *"Quantum materials are solids with exotic physical properties, arising from the quantum mechanical properties of their constituent electrons; such materials have great scientific and/or technological potential"*[30]. The simpler, rather empirical explanation that states *"I know one when I see it"* is perhaps more appealing to researchers in the field[31]. For the sake of clarity, throughout this perspective we regard any material in which emergent quantum phenomena is observed as a quantum material.

According to this rather inclusive explication, one can argue that the history of quantum materials is actually quite long. Various iron-based magnetic materials, for instance, have already been around for several centuries. Superconductivity won its first Nobel prize in 1913[32]. Topological states were experimentally realized via the integer quantum Hall effect in Silicon crystals in the 1980s[33,34]. This of course seeded the explosion of research for magnetic-field-free topological materials in the late 2000s[35-37]. Altermagnets, which exhibit both ferromagnetic and antiferromagnetic characteristics, have just recently joined the fray in 2020[38-40]. At any rate, it is clear that quantum materials have become increasingly more integrated into the scientific community.

Figure 1 provides a catalog of some of the hallmark properties of such quantum materials, as well as their various forms of preparation. When harnessed properly, it is evident that the fascinating electronic characteristics of these materials will have a transformative impact on both basic and pragmatic research alike. For example, topological materials with band structures that display spin-momentum locking offer atypical means to generate spin polarization in memory and logic devices[41-43]. Quantum spin liquids provide a rich experimental playground to fundamentally expand our understanding on many-body entanglement and fractionalized excitations[44,45]. Superconductors interfaced with topological insulators potentially lead to non-Abelian quasiparticles[46,47], which can be used to host qubit basis states that are inherently protected from decoherence[48,49].

The bright prospect of quantum matter is then quite apparent. Unfortunately, not all materials are inherently `quantum'. Depending on the material features that one seeks, a specific set of criteria may need to be satisfied in terms of composition and crystal structure. Recently, theoreticians have constructed a database that tabulates the majority of reliably-characterized crystal structures and screens their electronic properties for topological characteristics using first-principles calculations[50-52]. A specific set of design rules has also been proposed for altermagnetic materials[39]. Surprisingly, it turns out that in theory, such materials are more ubiquitous than expected, at least with regard to their mere existence.

Adequate experimental realization proves to be a completely different issue. Oftentimes the chemical composition, crystal phase, and dimensionality required by a quantum material are comparatively unconventional, and usually it is necessary to overcome a myriad of technical challenges for reasonable synthesis. As such, several different methods are utilized to grow quantum materials, and the geometry and size of the resulting samples also vary drastically, ranging from a handful of atoms all the way to bulk single crystals (Fig. 1)[53-57]. In other words, the common approach for quantum material synthesis is still somewhat exploratory, focused more toward probing new possibilities based on what is feasible in the context of sample preparation.

**The appeal of thin film quantum materials**

While all forms of quantum materials are useful for fundamental analyses or prototype demonstrations of functionality, thin films are especially more resourceful. We summarize their major strengths in Fig. 2. The capability to grow scalable material on standard substrates such as Si, GaAs, quartz, etc., widely implemented in the photonics and electronics industries provides an enormous advantage in the aspect of integrating a final device product onto a chip. This is well shown by the fact that metallic, semiconducting, and insulating thin films deposited in this way are already being regularly put to use even the most elaborate, modern device fabrication technologies[58-60].

Another perk of thin films is that they present the option to tune quantum material dimensionality. One-, two-, and three-dimensional (3D) crystals of the same material can exhibit strikingly disparate characteristics thanks to modifications in quantum confinement or the density of states. Provided that a robust thin-film deposition scheme is available, it is possible to manufacture materials with all such dimensions. Plain growth would result in simple 2D systems, while extended growth over long periods of time would yield material that has several-µm thickness, indistinguishable from its 3D bulk counterpart. Techniques such as cleaved-edge overgrowth[61-63], selective-area deposition[64-66], or even commonplace photolithography can be employed to synthesize 1D nanowires as well as 0D quantum dots from a thin-film basis.

Furthermore, thin film structures enable the utilization of heterogeneous interfaces. In certain cases, when two different materials are brought together, their collective band properties in conjunction with quantum confinement give rise to extraordinary outcomes. In fact, one can even argue that the recent enthusiasm toward topological quantum materials was sparked by the experimental realization of a quantum spin Hall insulator in $HgTe/Hg_xCd_{1-x}Te$ heterostructures[35-37]. More classic examples of quantum-matter heterostructures include 2D electron systems hosted in potential wells for the observation of electron-electron interactions[67-69].

**The status quo of thin film quantum material deposition techniques**

Given the benefits of thin film quantum materials, it is constructive to examine what tools can be used for their synthesis. Several approaches exist, such as molecular beam epitaxy (MBE), chemical vapor deposition (CVD), atomic layer deposition (ALD), pulsed laser deposition (PLD), sputtering or evaporation. Figure 3 depicts the characteristics of these techniques in terms of resultant sample purity and crystallinity. Throughout the rest of this section, we review each deposition method in detail, discussing their strengths and weaknesses, as well as studying representative examples of thin film quantum material growth.

- **Molecular beam epitaxy**

The innately fragile nature of most quantum states implies that electron scattering events can substantially shorten their lifetimes. Therefore, in order to properly observe and capitalize these exquisite features, thin film quantum materials with high crystallinity and minimized impurity concentrations are desired. While there are a number of techniques that can lead to such 'high-quality' thin film quantum materials, MBE is a strong leader.

The growth behavior of MBE thin films is governed by the interactions between the substrate surface and the beam flux of source material. Surface diffusion rates, sticking coefficients, adsorption lifetimes of the individual participating elements, as well as the reaction chemistries and thermodynamics at the growth front all play important roles. Consequently, the status of the initial growth surface is critical in determining the characteristics of MBE growth. The term 'epitaxy' is actually reserved for select cases where the crystallinity and texture of the deposited thin films track those of a single crystal substrate. Indeed, MBE was first developed in the late 1960s to meet the stringent demands of the microelectronics industry and form single-crystal thin films on semiconductor substrates [70-72].

Moreover, MBE has the extra benefit that the process entirely occurs in an ultra-high-vacuum (UHV) environment. This allows the growth of extremely pure material since unwanted impurity incorporation during deposition is minimized [73,74]. It is also possible to refine the elemental source materials under these UHV conditions which capacitates the growth of material that is cleaner than the original source itself[75,76]. Needless to say, over the past few decades researchers have often turned to MBE when in need of utmost sample quality. This is well demonstrated by the popularity of MBE-grown, ultra-high-quality GaAs/AlGaAs quantum well structures for the experimental study of many-body electron interactions, where near-perfect material systems are required [67-69, 74]. In fact, as shown in Fig. 4, these traits of MBE enabled the synthesis of GaAs/AlGaAs heterostructures that are the purest materials on earth[77]. In these systems, the defect concentration is calculated to be only 1 in 10 billion[77].

While the GaAs/AlGaAs system certainly serves as a landmark example, it is far from the only case in which MBE is the primary method for quantum material growth. Several metal oxide systems comprised of ZnO/MgZnO[78,79], $ABO_3$ perovskites [80-85], and cuprates have also been grown via MBE for the investigation of strongly-correlated electron systems[86-88]. With sufficiently high sample quality, the collective interactions of electrons within the involved band structure of transition-metal compounds emerge as enigmatic yet intriguing quantum phases of matter[89-91]. Other material groups, including pnictides[92-94], chalcogenides[68,95,96], as well as Heusler alloys are also being pursued in light of expanding the list of MBE thin film quantum materials [97-99].

Nevertheless, there are still some shortcomings of MBE that necessitate the search for alternative strategies for thin film quantum material synthesis. The effusion cells used to generate the molecular beams are typically based on thermal evaporation, and therefore it is rather difficult to achieve a reasonably high throughput for refractory metals with high melting temperatures. Metals that actively react with the source material crucibles or other chamber components also present a challenge. On top of that, constant overpressure of a constituent element is required during MBE growth to ensure facile stoichiometry control in target compound materials[100]. All in all, these constraints imply that despite its many strengths, not every quantum material can be grown readily using MBE.

### -Chemical vapor deposition

Unlike its physical deposition counterparts, CVD makes use of chemical precursors as its source material to grow thin films. Target metal elements are delivered to the growth chamber as gaseuous inorganic or organic compounds instead in the form of a pure melt or solid. At high enough substrate temperatures, the ligands that bind with the metal center in these chemical precursors decompose, leaving behind a thin film as a final reaction product. In this scheme, it is possible to combine several different precursors to deposit complex, multicomponent films. Additional reactant gasses such as $H_2O$, $O_2$, $O_3$, $NH_3$, $PH_3$, or $AsH_3$ can also be injected into the CVD chamber during growth to produce oxides, nitrides, phosphides, and arsenides[101,102].

Although CVD reactions can occur in ambient conditions, for the purpose of quantum material synthesis, it is desirable to perform the process in high vacuum. Under such controlled settings where almost all chemical reactions are dictated by molecules intentionally introduced to the CVD chamber, high-quality, epitaxial growth is achievable[103]. The constantly evacuated deposition environment also ensures that unintentional contaminants in the CVD-grown sample are significantly suppressed. Similar to MBE, these traits of vacuum-based CVD make it an attractive option for the preparation of thin film quantum materials. In fact, if we consider potential mass production, compared to MBE, CVD holds the additional advantage that it can readily scale processes to utilize 12-inch diameter wafers as well as batch schemes.

Unsurprisingly, CVD is a major player in the field. It excels particularly in areas where technical difficulties prevent MBE from offering high-quality sample growth, as highlighted in Fig. 5. For example, CVD is the method of choice for the synthesis of diamond, which is fairly difficult to grow using MBE due to the lack of suitable elemental source material[104,105]. Diamond has lately been studied extensively as a quantum system because it can act as a stable solid-state platform to investigate energetically deep defect states[106-108]. These 'color centers' mimic the properties of individual atoms, implying that quantum operations can be conducted by programming them in an adequate fashion[109-111]. Such artificial atoms can also be utilized as ultra-high-sensitivity quantum sensors[112-114].

Following the case of diamond, CVD is also prominent for the high-quality growth of other group IV materials. The melting temperatures of both Si and Ge are moderately high, and it is arduous to establish a sufficient beam flux for MBE. As a result, recent reports on quantum-grade Si/SiGe or Ge/SiGe heterostructures are predominantly based on CVD thin films[115-118]. These quantum material systems have been used to observe proximitized superconductivity and to host gate-defined spin and superconducting qubits[119-122].

One weakness of CVD as a thin film quantum material growth technique is the scarcity of commercially available precursors. Ideally, for fluent growth one wants liquid-phase chemical compounds that are stable, amply volatile, and cheap[123]. However, such options are rare even for materials commonly used in the semiconductor industry. Considering that quantum materials are typically composed of rather unconventional elements, it is unrealistic to expect that off-the-shelf precursor products are easily procurable. Also, since byproducts are inevitable in the CVD reaction process, sample purity can potentially become an issue in extremely high-end quantum applications.

**-Other deposition methods**

The advantages of MBE and CVD discussed in the previous paragraphs clearly demonstrate why they are preferable for the growth of thin film quantum materials. At the same time, it is important to contemplate the pros and cons of other frequently used thin film deposition methods. Depending on their attributes, there may be situations where they are potentially more applicable for a specific thin film quantum material system compared to MBE or CVD. Here, we evaluate the cases of PLD, sputtering, evaporation, and ALD.

In a PLD system, a high-power pulsed laser is focused onto the target source material to deposit thin film structures. In general, growth occurs in an ultra-high-vacuum environment just like MBE, which enables the synthesis of high-purity, epitaxial films under proper conditions. One major difference with MBE though is that the source material is flash vaporized in PLD. The laser beam generates a plasma plume of congruently evaporated species which, in theory, condenses stoichiometrically onto the substrate[124]. This aspect of PLD makes it explicitly useful for the growth of complex ternary and even quaternary thin film quantum materials with relatively precise compositions[125-128].

On the other hand, PLD also has some distinct drawbacks. In contrast to MBE, it is difficult to purify the source material used for deposition since melting does not occur universally. Impurity levels are therefore usually higher in PLD films compared to MBE films[91]. Also linked to the limited beam diameter of the laser, ablation of the target material occurs in a local fashion, which prohibits uniform thin film deposition over large substrate areas[129]. This implies the existence of a significant roadblock in realizing wafer-scale processes which are necessary for the fabrication of practical devices. Unwanted particulate formation on the growth surface is also a serious detriment that hinders the uniformity of PLD thin films[130].

Comparatively, sputtering, another popular physical vapor deposition technique for the growth of thin film quantum materials, has good process scalability and uniformity. During the sputtering process, electric-field-driven, ionized gas particles bombard the target material to eject atoms and deposit them onto a substrate[131]. While this intrinsically energetic source material vapor flux allows the rapid nucleation and growth of thin films, it simultaneously discourages the formation of large-area single crystals. Defects induced by structural damage to underlying layers can also have a negative effect on sample quality. Still, electronic properties that are suspected to derive from non-trivial topological band structures have been reported in various thin films deposited by sputtering[132,133].

In the context of quantum material growth, thermal or electron-beam evaporators can be thought of as higher-growth-rate, lower-vacuum MBE processes without the capability to monitor the surface conditions of the substrate. Both MBE and evaporation utilize source melts to grow films on a substrate, but high-quality, epitaxial growth is an anomaly for simple evaporation procedures. Consequently, evaporation is typically implemented for quantum material growth only when the benchmarks for crystallinity and purity are not too stringent. The deposition of metallic thin films for superconducting contacts or ferromagnetic applications are good examples[134,135].

Finally, we examine ALD as a prospective method to grow thin film quantum materials. A special subcategory of CVD, ALD is a step-by-step chemical process that is mediated by surface reactions[136]. The characteristic self-limited growth behavior inherent to ALD distinguishes it from other thin film deposition methods, which makes it possible to achieve conformal, nanometer-thick layers on ultra-high aspect ratio structures. For this reason, ALD is heavily utilized in the microelectronics industry where extremely sophisticated device fabrication schemes are necessary[58,59,136]. This suggests that if ALD thin film quantum materials are achievable, they have great potential for the production of novel functional devices since it would be possible to exploit such mature processing technologies.

At the time of writing this article, reports concerning stand-alone, ALD thin film quantum materials are still quite limited. Typically, layers grown by ALD have polycrystalline or amorphous microstructures, and the lack of long-range order in these samples can arguably prevent the manifestation of delicate quantum phenomena. That being said, several reports have demonstrated that the ALD growth of preferentially aligned crystalline films is possible[137-139]. Based on this fact, we speculate that ALD will soon gain more traction as a tool to synthesize thin film quantum materials. On a side note, ALD also shares the same issues with CVD in the context of precursor scarcity. To some extent, it is even worse because most of the time, ALD reactions are more complicated, further restricting the pool of selection.

Nonetheless, there have been a multitude of attempts to employ ALD thin films in quantum devices. For example, different types of ALD-grown, dielectric or metal thin films have been implemented as components of superconducting qubit circuits to enhance performance[135,140,141]. Diamond-based quantum sensors in which surface reactions are monitored via a thin ALD coating have also been reported[142,143]. In a broader perspective, these ALD films can also be classified as quantum materials, even though they may seem 'classic' on their own.

**Outlook**

There is still a long way to go for quantum materials to be utilized in our everyday lives. While several technical obstacles need to be overcome and new ones will certainly emerge, the capability to synthesize these exotic systems in thin film form will provide a robust foundation to build on. Just a few centuries ago, it would have been almost impossible to imagine the digital world as it is today, fulfilled by the invention of semiconductor-based field-effect transistors. What other game-changing, innovative functional devices will transform society as quantum material technologies progress? Will it be scalable qubit structures that initiate universal quantum supremacy? Maybe quantum sensors with unfathomable detection limits will let us scrutinize gravity waves and reveal long-lost secrets of the cosmos. Perhaps a new era of computing will come with the development of ultra-fast, non-volatile, and power efficient spin-orbit torque memory cells based on quantum matter. We can only speculate on what else, but one thing is for sure: thin film quantum materials will be at the heart of all of them.

**Acknowledgements**

This work was supported by the National Research Foundation of Korea(NRF) grant funded by the Korea government(MSIT)(RS202400348416)

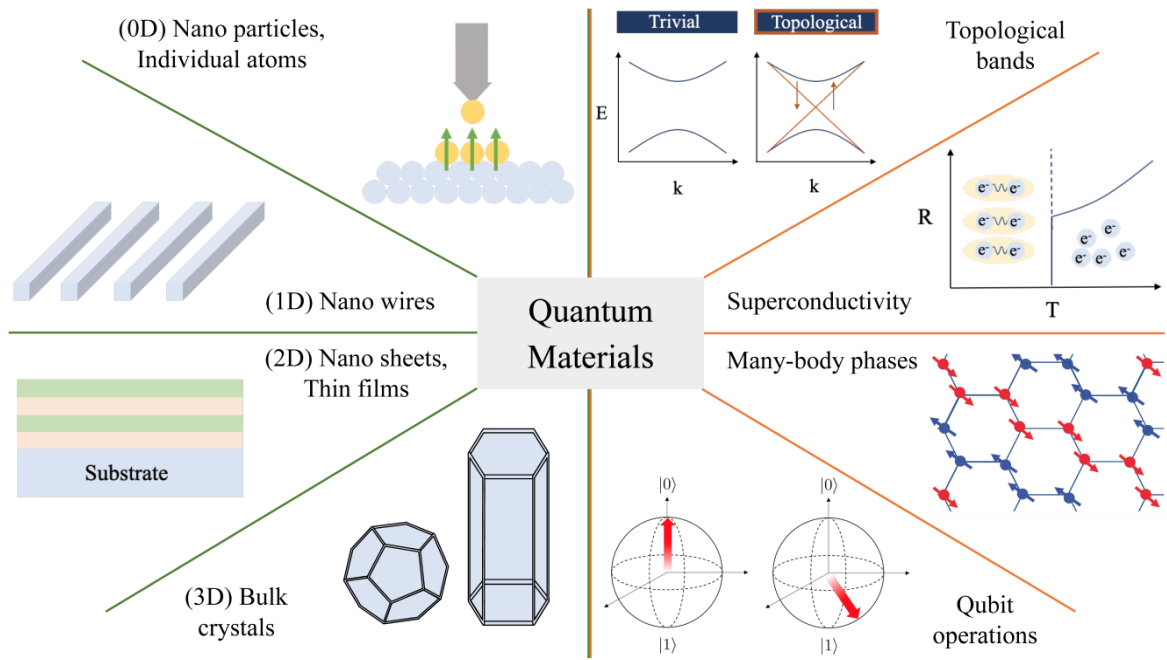

**Figure 1** Schematic examples of quantum material sample preparation and some of their hallmark features. Quantum materials have been reported across various different size scales and dimensionalities, starting from just a couple of atoms all the way to full bulk crystals. Characteristics of such quantum materials emerge in a number of distinct ways. Typical instances include topological bands, superconductivity, many-body electron phases, or two-level systems suitable for qubit operation.

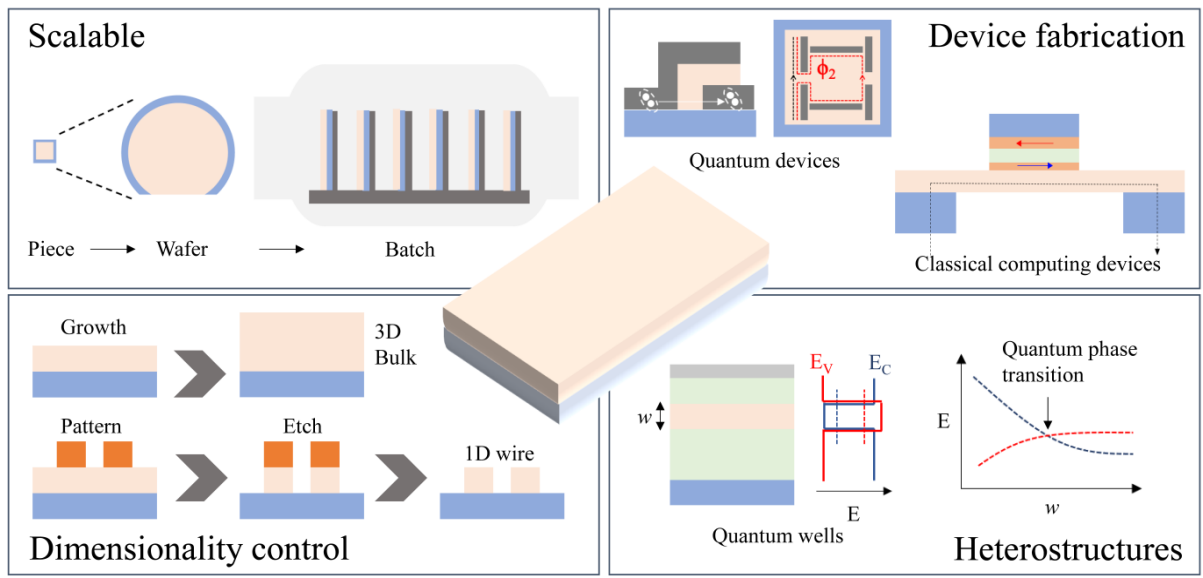

**Figure 2 The strengths of synthesizing quantum materials in the form of thin films.** Thin films are versatile since they grant access to all dimensionalities as well as opening up new possibilities through heterostructures. They also offer process scalability according to the size of the deposition substrate. The capability to fabricate complex functional devices such as Josephson junctions, interferometers, and spin-orbit torque memory structures is another plus.

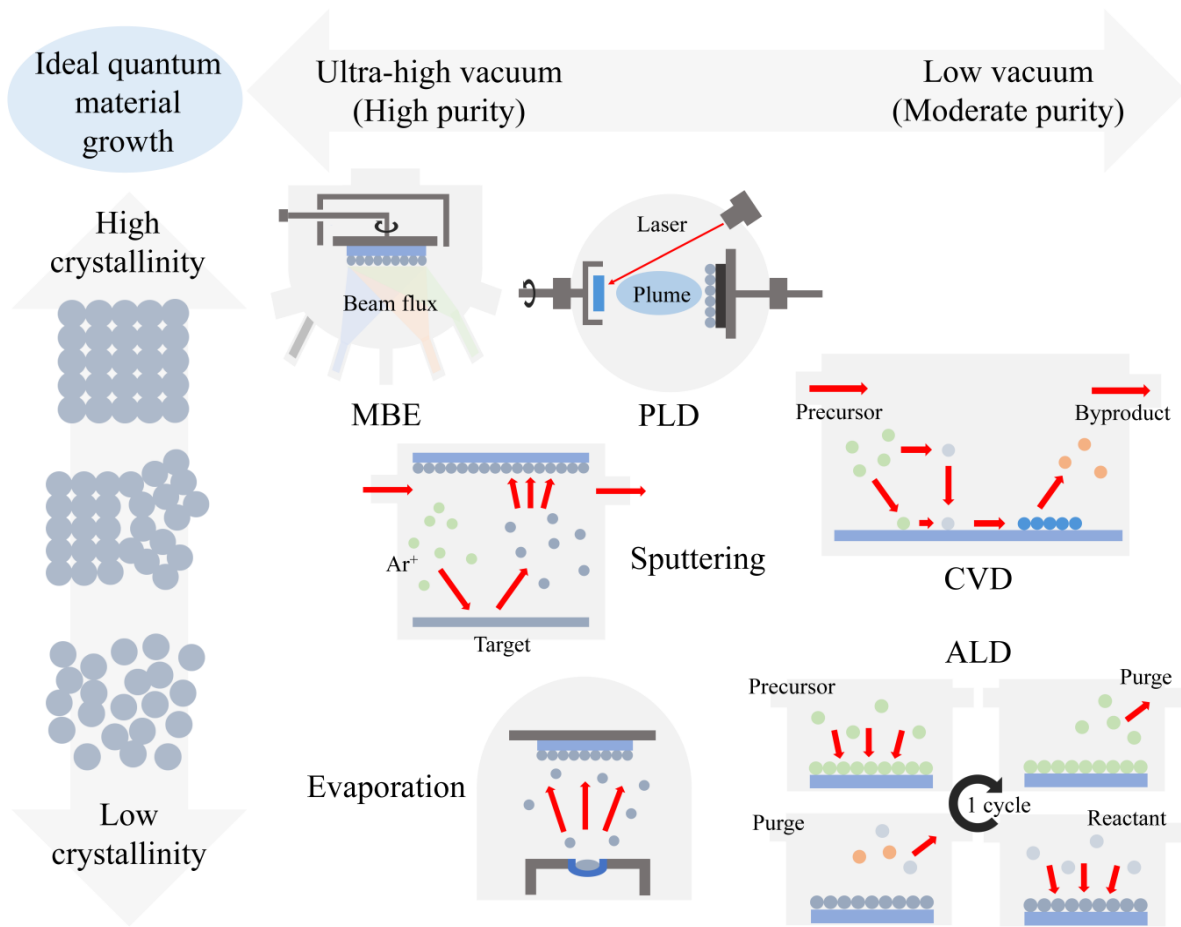

**Figure 3 Thin film quantum material growth techniques. All tools are placed on a spectrum with regard to typical crystallinity of resultant films and the expected sample purity based on vacuum conditions of the deposition chamber.**

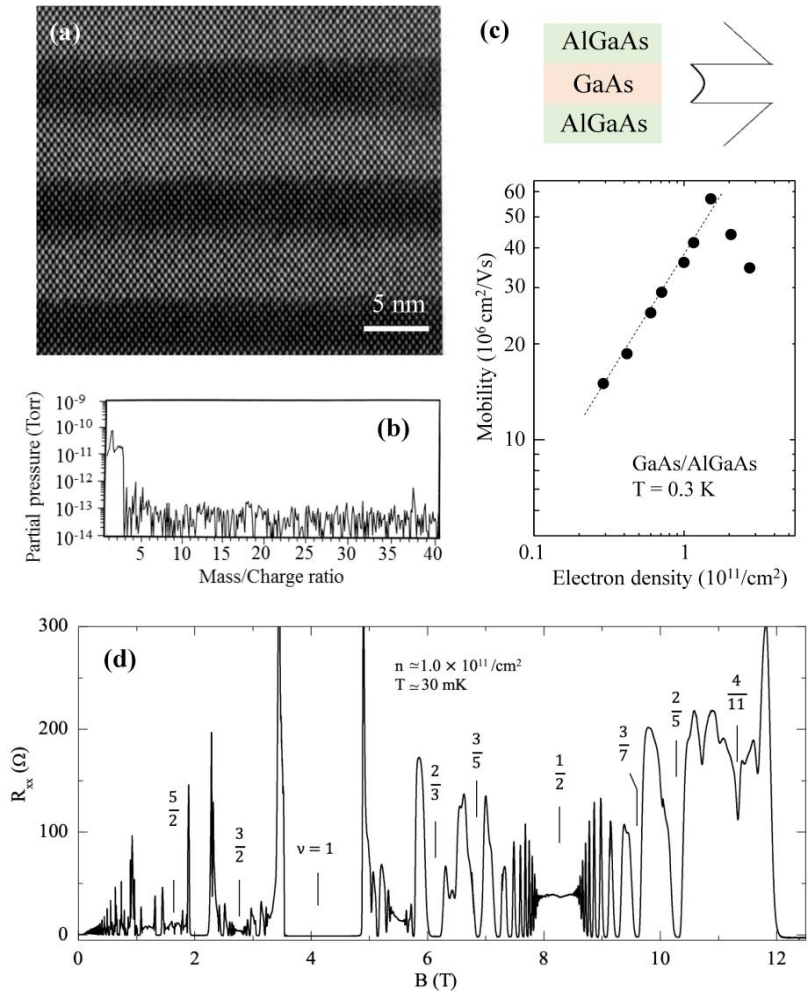

**Figure 4 MBE growth of quantum materials.** Single crystal, epitaxial superlattice structures can be grown using this technique as demonstrated by the high-resolution tunneling microscopy images shown in (a). The MBE chamber can be optimized so that virtually no impurities exist in the growth background, as demonstrated by the spectrum of an in-situ mass spectrometer (b). As a result, record-quality samples that display electron mobility values as high as ~6x10$^7$ cm$^2$/Vs can be grown. These samples have impurity levels on the order of 1 in 10 billion (c). In such systems, a plethora of many-body quantum phases can be observed in low-temperature magnetotransport measurements (d). Images adapted from Ref. 77.

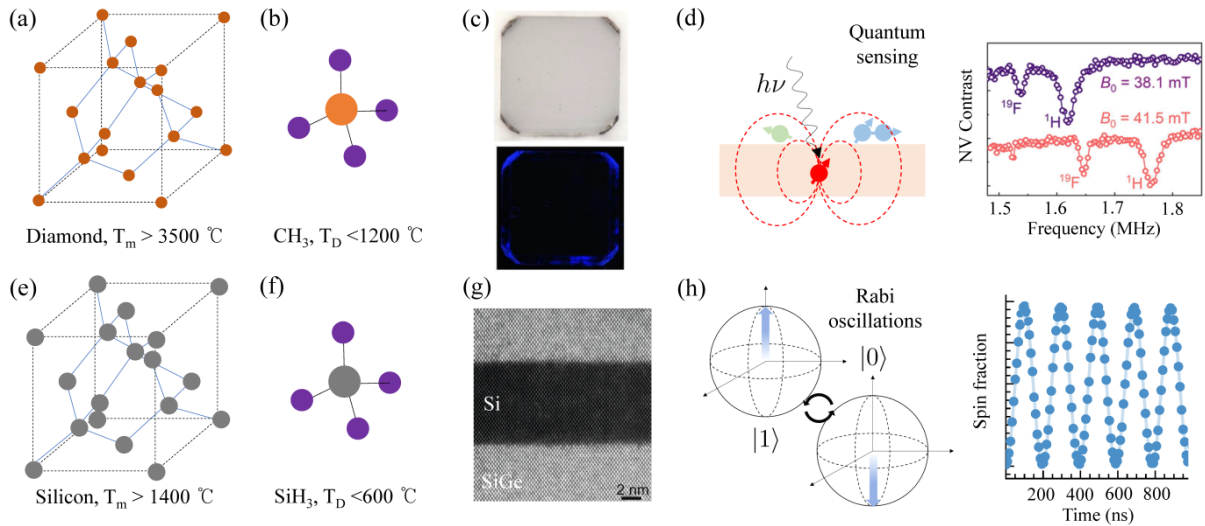

**Figure 5 Thin film quantum materials made from CVD.** In cases where the melting temperatures of elemental source materials are high ((a) Diamond, (e) Silicon), CVD provides an alternative route to high-quality thin film synthesis through chemical precursors with relatively lower decomposition temperatures ((b) $CH_3$, (f) $SiH_3$ for Diamond and Silicon, respectively). Films produced this way have excellent quality: (c) single-crystal diamond films with minimal defect photoluminescence (images adapted from Ref. 107) and (g) perfect epitaxial Si/SiGe interfaces (image adapted from Ref. 120). Applications such as quantum sensors (d) (data adapted from Ref. 112) and qubits (h) (data adapted from Ref. 121) are enabled by these high-quality CVD thin film quantum materials.